\documentclass[aip,amsmath,amssymb,reprint]{revtex4-1}
	\usepackage{graphicx}
	\usepackage{dcolumn}
	\usepackage{bm}
	\usepackage[utf8]{inputenc}
	\usepackage[T1]{fontenc}
	\usepackage{mathptmx}

\begin{document}
	\preprint{AIP/123-QED}
	\title{Multimode probing of superfluid $\mathbf{^4He}$ by tuning forks}
	\author{A.\,Guthrie}\email{a.guthrie1@lancaster.ac.uk}
	\author{R.\,P. Haley}
	\author{A.\,Jennings}
	\author{S.\,Kafanov}\email{sergey.kafanov@lancaster.ac.uk}
	\author{O.\,Kolosov}
	\author{M.\,Mucientes}
	\author{M.\,T.\,Noble}
	\author{Yu.\,A.\,Pashkin}
	\author{G.\,R.\,Pickett}
	\author{V.\,Tsepelin}
	\author{D.\,E.\,Zmeev}
		\affiliation{Department of Physics, Lancaster University, Lancaster, LA1 4YB, United Kingdom}
	
	\author{V.\,Efimov}
		\affiliation{Institute of Solid State Physics RAS, Chernogolovka, 142432, Moscow Region, Russia}
	
	\begin{abstract}
		Flexural mode vibrations of miniature piezoelectric tuning forks (TF) are known to be highly sensitive to superfluid excitations and quantum turbulence in $\mathrm{^3He}$ and $\mathrm{^4He}$ quantum fluids, as well as to the elastic properties of solid $\mathrm{^4He}$, complementing studies by large scale torsional resonators. Here we explore the sensitivity of a TF, capable of simultaneously operating in both the flexural and torsional modes, to excitations in the normal and superfluid $\mathrm{^4He}$. The torsional mode is predominantly sensitive to shear forces at the sensor - fluid interface and much less sensitive to changes in the density of the surrounding fluid when compared to the flexural mode. Although we did not reach the critical velocity for quantum turbulence onset in the torsional mode, due to its order of magnitude higher frequency and increased acoustic damping, the torsional mode was directly sensitive to fluid excitations, linked to quantum turbulence created by the flexural mode. The combination of two dissimilar modes in a single TF sensor can provide a means to study the details of elementary excitations in quantum liquids, and at interfaces between solids and quantum fluid.
	\end{abstract}
	\maketitle
	\label{intro}
		Piezoelectric tuning forks (TF) have found increasing use in scientific applications, particularly in fluid engineering \cite{toledo2014application}, due to their compact size and high sensitivity to liquid viscosity. In fluid mechanics involving quantum media TFs have been used for measuring a variety of phenomena including quantum turbulence \cite{blavzkova2007transition,sheshin2008characteristics,bradley2009transition,blavzkova2009generation,schmoranzer2019dynamical}, acoustic (first sound) emission \cite{bradley2012crossover,salmela2011acoustic,schmoranzer2011acoustic}, cavitation \cite{blavzkova2008cavitation,blavzkova2008cavitation1}, Andreev scattering \cite{bradley2008probing,bradley2009damping,vclovevcko2011high}, NMR excitation \cite{clovecko2019NMR}, and quasiparticle detection in superfluid $\mathrm{^3He}$ \cite{bradley2017andreev}. Previous low temperature experiments typically utilise the flexural mode of a TF, such that the tines move inwards and outwards in anti-phase with each other. The fundamental flexural mode frequency is typically a few tens of $\mathrm{kHz}$, with some experiments also using higher harmonics \cite{bradley2012crossover}. Flexural motion interacts with a medium by displacing the adjacent fluid and generating both normal and shear stress. A torsional mode (where the tines twist in opposite directions) creates predominantly shear stress at the moving interface. In the past, large torsional oscillators have been used for measurements of the normal-fluid fraction below the superfluid transition temperature in $\mathrm{^4He}$ \cite{andronikashvili1948andronikashvili}, however, properties of quantum fluids, particularly vorticity generation, have yet to be studied under high frequency shear stress.

		In many experiments involving oscillators, the device performs the actuation and detection of the phenomena under study. The measured response is then a combination of the effects of generation and detection. Conventional quantum turbulence measurements using oscillating wires and TFs provide an example of this, where the velocity is increased until a characteristic increase in damping is observed \cite{bradley2005turbulence}. The velocities at which these events occur are known as critical velocities. The advantage of this scheme is the local nature of the probe; vortices created by the oscillator are measured in the same location they are created and propagation of vortices need not be considered. The disadvantage of this approach is the use of the same mode for detection. The oscillator response must first be measured in vacuum to eliminate effects that are not a result of the medium under study.	Conversely, experiments involving separate emitter and detector probes inevitably introduce some spatial separation and must account for propagation mechanisms when modelling the system \cite{nago2011time}. 
	
	\begin{figure}[b]
		\centering    
		\includegraphics[width=\linewidth]{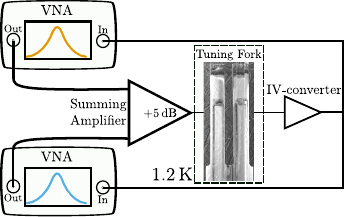}
	    \caption{Schematic showing an experimental setup for measuring two modes on the same tuning fork. Two vector network analyzers (VNAs) provide AC driving voltages at two frequencies coming to the inputs of a summing amplifier with $5\,\mathrm{dB}$ gain, also used to increase the dynamic range of the setup. The output current is passed through an I-V converter with $10^3 \, \mathrm{V\, A^{-1}}$ gain  \cite{holt2012electrometric}, and the two signals recovered using a splitter before being measured at each frequency by the two VNAs. A scanning-electron-microscope (SEM) image of the TF is shown as part of the schematic.}
	    \label{setup1}
	\end{figure}	
	
	Here we present an approach using two vibrational modes on the same TF, using one mode to create vorticity, while probing via the second mode. No vacuum response calibration is needed since we are monitoring changes in absolute damping, rather than deviations from calibration data as we vary excitation on the first mode. We use this technique to probe superfluid $\mathrm{^4He}$, demonstrating the sensitivity of the torsional mode to fluid perturbations created by the flexural mode, which may include quantum turbulence. 

	The torsional modes are standing, rotational waves in the TF tines, with one end clamped at the base, the other end free. Torsional waves are not dispersive, meaning that their propagation velocity does not depend on the oscillation frequency. In addition, the twisting angle, $\theta (x)$, and correspondingly the tangential velocity of the fundamental mode depends linearly on the distance from the base. This leads to a more uniform interaction of the TF with the medium when compared to the flexural mode where the tip velocity makes the largest contribution.

	The fundamental frequency $f_0$ of torsional motion is determined by the dimensions and mechanical properties of the tine as \cite{song2006coupling}
	\begin{equation} \label{tors:freq}
	    f_0 = \frac{1}{2l}\sqrt{\frac{3G}{\rho}\frac{\beta  w^2}{t^2 + w^2}},
	\end{equation}
	where $l$, $t$ and $w$ are the length, thickness and width of the TF tines, respectively, $\rho = 2650 \, \mathrm{kg \, m^{-3}}$ is the quartz density, and  $G=3.11 \times 10^{10} \, \mathrm{N m^{-2}}$ is the shear modulus of quartz. The dimensionless coefficient, $\beta$, depends on the ratio of $w/t$  changing from $\beta = 0.141$ for $t=w$ to $0.333$ for $t \ll w$ \cite{song2006coupling}.

	When considering measurement of a TF it is important to convert measured electrical signals to the mechanical parameters which define the system. For flexural oscillations there are well established relationships between the force, $F$, and the excitation voltage, $V_0$, as well as between the velocity, $v$, and measured current, $I$, via the flexural fork constant, $a_f$ \cite{Blaauwgeers2007quartz,karrai2000lecture,bradley2012crossover}. By analogy, the torsional fork constant, $a_{t}$, can be defined for torsional motion where the torque, $\tau$, is given by, 
	\begin{equation}
	    \tau = a_t V_0 /2.
	\end{equation}
	The work done per unit time is now $\sim \tau \omega_t$, where $\omega_t$ is the angular velocity, with the power dissipated in the fork ($\sim IV_0$). The measured current, $I$, is expressed as
	\begin{equation}
	    I = a_t \omega_t.
	\end{equation}

	The devices used in this experiment are quartz tuning forks with tine length $1.61\,\mathrm{mm}$, width $0.22\,\mathrm{mm}$ and thickness $0.14\,\mathrm{mm}$, with the electrode geometry allowing effective excitation of the torsional mode \cite{kawashima1996quartz}. The devices have a well defined torsional oscillation mode at $393\,\mathrm{kHz}$ and a flexural mode at $76\,\mathrm{kHz}$ in vacuum. The torsional frequency is close to the calculated value of $400\,\mathrm{kHz}$ using Eq.\,\ref{tors:freq} with $\beta=0.20$. The devices are driven piezoelectrically, using an AC voltage supplied by the output port of a vector network analyzer (VNA) and the response recorded by the input port of the VNA as a function of frequency, as shown in Fig.\,\ref{setup1}. In this scheme, the two VNAs can be operated either individually, or together, to achieve excitation and detection at a single frequency, or simultaneous excitation and detection at both frequencies. The resultant curve is fitted to give the frequency, resonance width and amplitude. A $5 \,\mathrm{dB}$ summing amplifier was used to increase the range of drive forces we could supply to the forks.

	The measurement setup is calibrated to account for any impedance mismatch in the experimental circuit by replacing the TF with a $38\,\mathrm{k\Omega}$ resistor and setting the value of current to $0\,\mathrm{dB}$ on the VNA. The velocity was then quantified by comparison to displacement measured by optical methods. These were performed using an LDV Polytec OFV-2570 laser-Doppler-vibrometer to record the tip linear velocity as a function of driving voltage, a well established technique for TF calibration \cite{gao2015research}. Electrical and optical measurements have been previously demonstrated to agree within $10\%$ \cite{efimov2009direct,ahlstrom2014frequency} of each other for flexural oscillations. This allowed the fork constant for the flexural mode to be determined as $a_f= 2.81 \times 10^{-6} \, \mathrm{C \, m^{-1}}$. For the torsional mode, the linear velocity projections recorded by the Dopper-vibrometer were converted to an angular velocity using an effective radius, $r = \sqrt{t^2 + w^2} /2$, yielding a fork constant $a_t= 7.51 \times 10^{-10} \, \mathrm{C \, \mathrm{rad}^{-1}}$. The velocities presented in this paper are calculated using this calibration technique.
	
	The experiments described in this paper were performed in an immersion cryostat with a base temperature of $\mathrm{1.2\,K}$. The helium temperature was inferred from $^4$He saturated vapour pressure \cite{donnely1998observed}. The forks were mounted onto a PCB with copper traces, connected using high-frequency SMP and SMA feedthroughs to the room temperature electronics. The assembly was attached to the end of a brass probe submerged directly in the helium bath. 

	\begin{figure}[t] 
		\includegraphics[clip,width= \columnwidth]{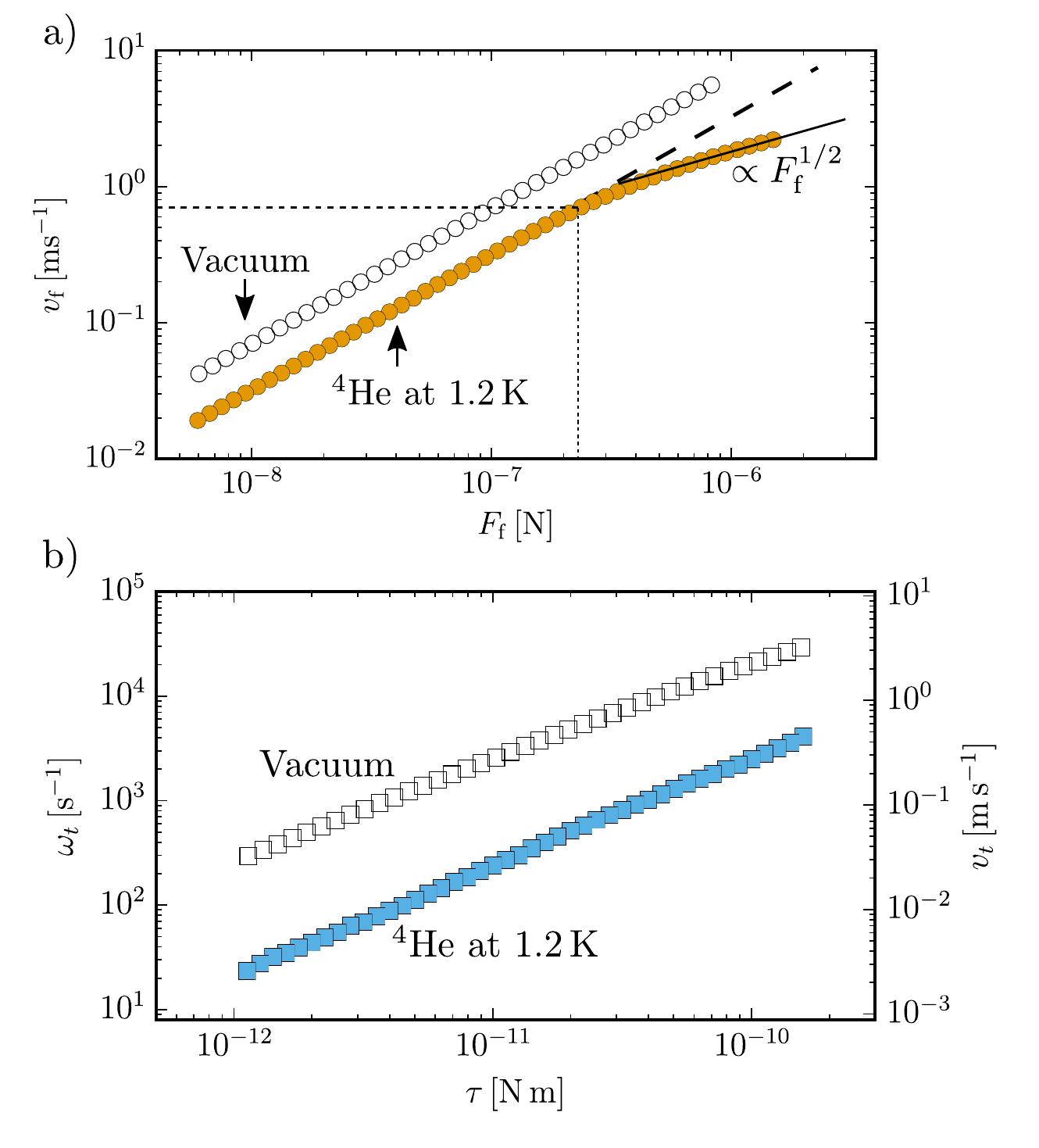}
		\vspace{-5mm}
		\caption{(a) \label{Fig2} Peak tine tip velocity, $v_\mathrm{f}$, as a function of driving force, $F_\mathrm{f}$, for the flexural mode of a TF in vacuum and $\mathrm{^4He}$ at $1.2 \, \mathrm{K}$. In vacuum the TF demonstrates linear behaviour with the damping dominated by internal losses. In $\mathrm{^4He}$ at $1.2 \, \mathrm{K}$ the losses are higher due to normal-fluid damping in the condensate. The transition to the turbulent regime is demonstrated by the change in gradient occurring at $70 \, \mathrm{cm \, s^{-1}}$, as indicated by the black dotted lines. The solid black line serves as a guide for the eye of the expected asymptotic dependence in this regime, while the dashed black line indicates the continued asymptotic dependence of the linear regime. (b) Peak tine angular velocity as a function of driving torque for the torsional mode of a TF in vacuum and $\mathrm{^4He}$ at $1.2 \, \mathrm{K}$. The high damping in $\mathrm{^4He}$ is a result of strong acoustical (first-sound) emission, due to the high frequency of this mode. No turbulent transition is observed up to an angular velocity of $5 \times 10^3 \, \mathrm{rad \, s^{-1}}$, corresponding to $\sim 50 \, \mathrm{cm\, s^{-1}}$.}
	\end{figure}

	Figures\,\ref{Fig2}(a) and \ref{Fig2}(b) present the force-velocity relationships for the flexural and torsional mode, in vacuum at $300 \, \mathrm{K}$ and liquid $\mathrm{^4He}$ at $1.2 \, \mathrm{K}$. In vacuum, the flexural mode demonstrated a linear relationship between force and velocity over a large range. The torsional mode remained linear at low velocities, however, demonstrated some weak non-linear behaviour at high velocities, ($>1\,\mathrm{m\,s^{-1}}$), occurring at a torque of $3 \times 10^{-11} \, \mathrm{N \, m}$, corresponding to an angular velocity of $\sim 10^4 \, \mathrm{s^{-1}}$. The measurements in $\mathrm{^4He}$ were performed well below the range of the vacuum non-linearity. Measurements in vacuum gave $Q$-factors for the flexural and torsional modes of $4.7 \times 10^4$ and $4.8 \times 10^4$, respectively. In superfluid $^4$He at $1.2 \, \mathrm{K}$ these $Q$-factors dropped to $1.8 \times 10^4$ and $5.5 \times 10^3$. The damping increase for the flexural mode can be explained by Stokes' drag resulting from the normal-fluid fraction, however, the order-of-magnitude increase of damping for the torsional mode we attribute to the presence of significant acoustic emission. 

	High frequency oscillators are known to emit sound waves in $\mathrm{^4He}$. The emission for flexural motion of TFs follows a quadrupole model with resonance width, $\Delta f \propto f_0^5$, where $f_0$ is the resonance frequency \cite{bradley2012crossover}. The corresponding relationship for the torsional mode is unknown. However, the geometry suggests we should expect the damping to follow an octupole emission model since each twisting tine will act as an acoustic quadrupole. We therefore expect a higher order dependence for torsional oscillators compared to flexural ones, evidence by the significant acoustic damping observed on the torsional mode in $\mathrm{^4He}$. In addition to the damping increase, sidebands were observed on the frequency response for the torsional mode, an effect we attribute to acoustic resonances in the experimental volume. Reducing the high levels of acoustic emission could be achieved by developing lower frequency torsional modes, through the use of longer forks or more flexible materials.

	For higher velocities in $^4$He, we must consider the effects of quantum vorticity on the TF. In a quantum fluid, much like a classical fluid, there are two flow regimes defined by the Reynolds number \cite{landau1987fluids,ahlstrom2014frequency}
	\begin{equation} 
			\mathrm{Re} = \frac{\rho_\mathrm{He} v L}{\mu},
	\end{equation}
	where $\rho_\mathrm{He}$ is the fluid density, $v$ is the tip velocity of the fork relative to the fluid, and $\mu$ is the dynamic viscosity. For low Reynolds numbers ($\mathrm{Re} \lesssim 1$) we expect laminar flow while for high Reynolds numbers ($\mathrm{Re} \gg 1$) the flow should be turbulent. Turbulent flow is not exactly solvable within the framework of Navier-Stokes, however, it can be empirically represented by a drag coefficient, $C_D$, with the drag force
	\begin{equation} \label{eq:turb}
			F_T = -\frac{1}{2} C_D A \rho_\mathrm{He} v^2,
 	\end{equation}
	where $A$ is the cross-sectional area of the plane perpendicular to the flow. From this, the turbulent regime is characterized by $v \propto F_T^{1/2}$. The crossover between these regimes occurs at a critical velocity given by \cite{ahlstrom2014frequency}
	\begin{equation} \label{criticalvel}
		v_c = \sqrt{\gamma \omega \kappa},
	\end{equation}
	where $\gamma \sim 1.7$ is a geometrical constant and $\kappa = h/m$ is the constant of circulation, with $h$ and $m$ being Planck's constant and $^4$He atomic mass, respectively. 

	Figure\,\ref{Fig2}(a) shows the transition to the turbulent regime on the flexural mode at a velocity of $\sim 70 \, \mathrm{cm \, s^{-1}}$. This is characterized by the ``kink'' in the data, transitioning from a linear dependence to a regime of higher damping.  Previous work has verified the appearance of a 'kink' as the turbulent transition by using measurements of second sound to probe vortex-line density  \cite{jackson2016measurements}. The measured critical velocity is consistent with that measured in previous turbulence experiments involving tuning forks \cite{ahlstrom2014frequency}, and with that given by Eq.\,\ref{criticalvel}. 

	Despite moving with an angular velocity of $5 \times 10^3 \, \mathrm{rad \, s^{-1}}$, corresponding to a linear velocity of the tip $\sim 50 \, \mathrm{cm\,s^{-1}}$, the torsional mode had no obvious transition to a turbulent regime, as shown in Fig.\,\ref{Fig2}(b). Given the order-of-magnitude difference in the resonance frequencies, we expect the critical velocity for the torsional mode to be considerably higher than that of the flexural mode.  The dependence given by Eq.\,\ref{criticalvel} would suggest that the critical velocity for turbulent onset for the torsional motion of this TF may be close to $10^{4}\,\mathrm{rad \, s^{-1}}$. The higher acoustic damping of the torsional mode, combined with an increased critical velocity, made it difficult to achieve the velocities required for turbulence creation. Nevertheless, the linear response of the torsional fork over a large range makes this mode ideal as a detector for quantum fluid excitations produced by another mode. 

\begin{figure}[t]
		\includegraphics[width=\linewidth]{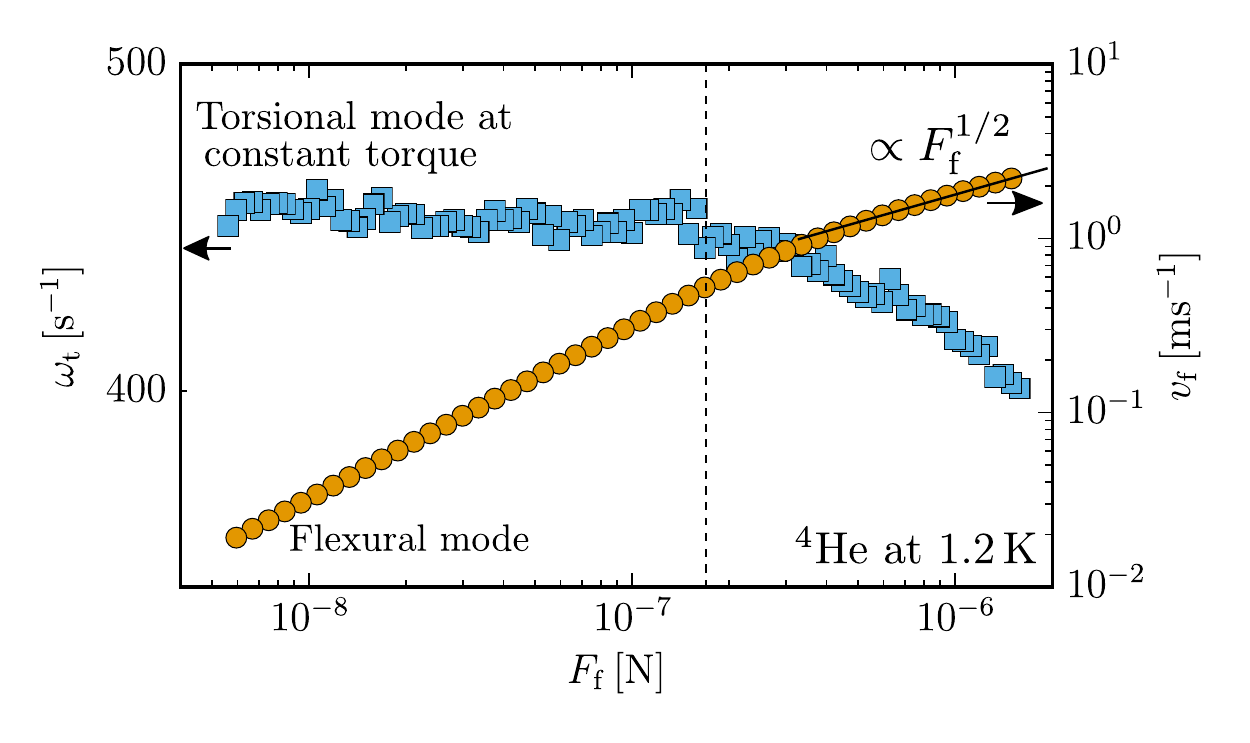}
		\vspace{-10mm}
		\caption{\label{twomodes} Force-velocity relationship for the coupled, two-mode detection scheme. The orange circles show the force response for the $76\,\mathrm{kHz}$ flexural mode in bulk $\mathrm{^4He}$ at $1.2\,\mathrm{K}$. A clear transition to a turbulent regime is observed where the slope, representing the damping, increases. The increase occurs at the critical velocity, $v_c$, as indicated by the dashed black line. The solid black line serves as a guide for the eye of the expected asymptotic dependence. The blue squares show the velocity of the torsional mode as a function of the force on the flexural mode. At the same value of force corresponding to the critical velocity on the flexural mode a transition to higher damping occurs on the torsional mode indicating that the torsional mode is sensitive to vorticity generated by the flexural mode.}  
	\end{figure}

	Having two well defined modes with a large frequency separation opens the possibility for separating the driving and detection between the modes. The setup shown in Fig.\,\ref{setup1} was used as a detection scheme for excitations related to quantum turbulence produced in superfluid $\mathrm{^4He}$ at $1.2\,\mathrm{K}$. Figure\,\ref{twomodes} shows the force-velocity relationship for the flexural mode at $\mathrm{^4He}$ at $1.2\,\mathrm{K}$, alongside the velocity of the torsional mode measured simultaneously. Here, the drive on the torsional mode is kept constant at a peak torque of $2 \times 10^{-11} \, \mathrm{N \, m}$, and the force, $F_\mathrm{f}$, refers to the force being applied to the flexural mode.

	Remarkably, monitoring the torsional mode at constant drive whilst increaseing the flexural drive demonstrated that the torsional mode was sensitive to excitations brought about by the transition to turbulence in the flexural mode. The transition to higher damping is evidenced by the shift to a decreasing angular velocity for the same drive force as the amount of drag on the tuning fork increases. Importantly, this transition occurs at the same force corresponding to the onset of turbulence in the flexural mode. Crosstalk can be ruled out since the resonance width of both modes is far less than the frequency spacing between them. In addition, at $1.2\,\mathrm{K}$ there is no significant contribution from the acoustic sidebands.

	The detection technique presented has the advantage that it does not rely on a previous calibration of the detection mode in vacuum. However, it is not applicable at very high drives due to a transition away from simple harmonic oscillator dynamics. In this regime higher-order mixing terms between the different resonant modes become non-negligible, limiting this type of experiment. 

	In conclusion, the initial measurements of torsional tuning forks in bulk liquid $\mathrm{^4He}$ have been performed down to $1.2\,\mathrm{K}$. The torsional mode experienced high damping in liquid compared to vacuum as a result of acoustic emission. This is supported by the presence of sidebands in the frequency response, a result of acoustic resonances in the experimental space. It has been previously shown that high oscillation frequencies lead to high critical velocities for the onset of turbulence; the turbulent transition was not observed in the torsional mode up to an angular velocity of $5 \times 10^3 \, \mathrm{rad \, s^{-1}}$. The absence of a turbulent onset on the torsional mode allowed us to probe the excitations created by the flexural mode without the introduction of significant spatial separation.

	The torsional mode damping increased at the same point at which the flexural mode transitioned to a turbulent regime, suggesting that the detection mode is sensitive to excitation related to the vorticity from the generator mode. The sensitivity could be improved through the use of higher $Q$ torsional oscillators and by reducing the high levels of acoustic emission. The work presented here paves the way towards the use of higher frequency torsional oscillators as low-temperature probes, particularly as sensitive detectors of localised topological defects. Our results also highlight the limitation of using high frequency devices as probes in liquid $\mathrm{^4He}$ due to substantial acoustic emission.

	We acknowledge the excellent technical support of A.~Stokes, M.\,G.~Ward. We thank S. Autti,  M.~Poole, R.~Schanen and A.\,A. Soldatov for support in the measurements and useful scientific discussions. This research was supported by the UK EPSRC Grants No. EP/L000016/1, EP/P024203/1, EP/K023373/1 and No. EP/I028285/1, and the European Microkelvin Platform, ERC 824109. All data used in this paper are available at http://dx.doi.org/10.17635/lancaster/researchdata/xxxx, including descriptions of the data sets.

\end{document}